\documentclass[3p]{elsarticle}

\makeatletter

\usepackage{color}
\usepackage{amssymb}
\usepackage{amsmath}
\usepackage{lineno}

\begin{document}
\begin{frontmatter}

\title{Evolution of trust in a hierarchical population with punishing investors}
\author[add1]{Ketian Sun}
\author[add1]{Yang Liu}
\author[add1]{Xiaojie Chen\corref{correspondingauthor1}}
\cortext[correspondingauthor1]{Corresponding author: xiaojiechen@uestc.edu.cn}
\author[add2]{Attila Szolnoki}

\address[add1]{ School of Mathematical Sciences, University of Electronic Science and Technology of China, Chengdu 611731, China}

\address[add2]{Institute of Technical Physics and Materials Science, Centre for Energy Research, P.O. Box 49, H-1525 Budapest, Hungary}

\begin{abstract}
Trust plays an essential role in the development of human society. According to the standard trust game, an investor decides whether to keep or transfer a certain portion of initial stake to a trustee. In the latter case, the stake is enhanced to signal the value of trust. The trustee then chooses how much to return to the investor. We here distinguish two types of investors and two types of trustees who can learn from each other. While a trustee can be trustworthy or untrustworthy, an investor could be normal or punishing one. The latter strategy punishes both untrustworthy trustees and normal investors who are reluctant to control misbehaving trustees. Importantly, we assume a hierarchical population where the portion of investors and trustees is fixed. By means of replicator equation approach, we study the $N$-player trust game and calculate the level of trust and trustworthiness. We find that the introduction of punishment can induce a stable coexistence state between punishing investors and trustworthy trustees. Furthermore, an intermediate fraction of investors can better promote the evolution of trust when the punishment intensity is low. For more intensive punishment, however, a higher fraction of investors can be more efficient to elevate the trust level. In addition, we reveal that appropriate increase of the punishment intensity can enlarge the attraction domain of the coexistence state.
\end{abstract}
		
\end{frontmatter}

\section{Introduction}
\label{}
\indent
Trust has a fundamental impact on how human society evolves \cite{manapat2013information,chica2019effects,charness2011trust,fu2019trust,fehr2009economics,chica2017networked,koster2013opening,pinyol2013computational,abbass2016review,abbass2016trusted,johnson2011trust}. Importantly, with the help of trust, interactions between individuals can be simplified, which is conducive to social development. For instance, it has been found that higher levels of trust have been related to more efficient judicial system, higher quality government, and greater financial development \cite{guiso2004role}.

Therefore it is vital to find and identify those mechanisms which can enhance and maintain trust in a population of rational individuals. The common paradigm is the so-called trust game (TG) \cite{berg1995trust}, which has been studied extensively in the last decade \cite{tarnita2015fairness,kvaloy2017trust,yi2021game,abbass2015n,kumar2020evolution,lim2020stochastic,fang2021evolutionary}. For example, Abbass {\it et al.} proposed the $N$-player trust game (NTG) in an infinite well-mixed population, and stressed that even a small initial portion of untrustworthy individuals could be detrimental because the population evolves to a state with zero truster and many untrustworthy individuals \cite{abbass2015n}. Furthermore, Kumar {\it et al.} assumed a scale-free interaction network with degree non-normalized dynamics and found that there are parameter values for which both trust and trustworthiness can survive \cite{kumar2020evolution}. Subsequently, Lim introduced a two-population model of the TG with asymmetrical features, and proved that stochastic evolutionary dynamics with asymmetrical factors can lead to the evolution of high trust level with high trustworthiness \cite{lim2020stochastic}. Recently, Fang {\it et al.} showed that the consideration of reward and punishment can induce the stable coexistence state of investors and trustworthy trustees in the NTG, which indicates that the evolution of trust can be greatly promoted \cite{fang2021evolutionary}.
	
It is important to stress that populations often have hierarchical structures in which groups are not necessarily equally large \cite{lim2020stochastic,deng2020cooperation,suarez2021reciprocity,nockur2021different,foley2021avoiding,przepiorka2021emergence,liu2021effects,hauser2019social,wang2010effects,otten2020heterogeneous}. However, just a few studies have considered the asymmetrical nature of the NTG, which implies the differences in roles between investors and trustees.
Therefore it is theoretically challenging to reveal how trust evolves in a population where these groups may cover different portions of the whole population.
	
To explore this research avenue, we here consider the NTG in an infinite population where players in both groups have two options. In particular, similar to the standard approach, we assume that a trustee can choose to be a trustworthy individual who returns investment or to be an untrustworthy individual who does not return any part of the received investment. Furthermore, inspired by previous works about the efficiency of punishment to keep cooperation level high \cite{boyd2003evolution,dreber2008winners,egas2008economics,szolnoki2013correlation,szolnoki2017second,Perc2017PR,huang2018evolution,Wang2018AMC,chen2018punishment,calabuig2016carry,capraro2021punishing}, we assume that investors could be normal or punishing. In particular, the punishing strategy will spend some budgets to punish both untrustworthy trustees and normal investors who refuse to impose the mentioned punishment. By means of replicator equations, we show that the introduction of punishment can lead to the stable coexistence state of punishing investors and trustworthy trustees. Due to the hierarchical population, the specific portions of two main groups have crucial importance. Namely, by means of numerical calculations, we reveal that for low punishment intensity, trust evolves better when the fraction of investors is intermediate. For high punishment intensity, however, high trust level can be reached when the fraction of investors is high. We further find that reasonable increase of the punishment intensity is beneficial to the evolution of trust. In this work, we accordingly establish a theoretical model for studying the evolutionary dynamics of trust in a hierarchical population with punishing investors for the first time and further reveal the effects of punishment on the evolution of trust in the scenario of hierarchical populations from a theoretical perspective.

The remaining of this paper is structured as follows. In the next section we introduce our NTG model in a hierarchical population with punishing investors. This is followed by the presentation of our theoretical analysis and numerical calculations in Section~3. Finally, we draw our conclusion in Section~4. Notably, the details of our analysis are given in an extensive Appendix.
	
\section{Model}

We consider an infinite well-mixed population where we have two groups with fixed sizes: they are the investor community and the trustee community. Importantly, since their sizes could be different, we assume that the fraction of investors in the whole population is an $\alpha$ ($0<\alpha<1$) constant, while the fraction of trustees is $1-\alpha$. At each time step, $N$ ($N>2$) players are chosen from the two communities to form an interacting group and participate in a trust game. For each investor, there are two strategies to be chosen: being a normal investor (strategy $M$), or being a punishing investor (strategy $P$).
In parallel, each trustee player there are also two options: being a trustworthy trustee (marked by strategy $T$) or an untrustworthy trustee (called strategy $U$).
	
\begin{figure}[h!]
	\centering
	\includegraphics[width=9cm]{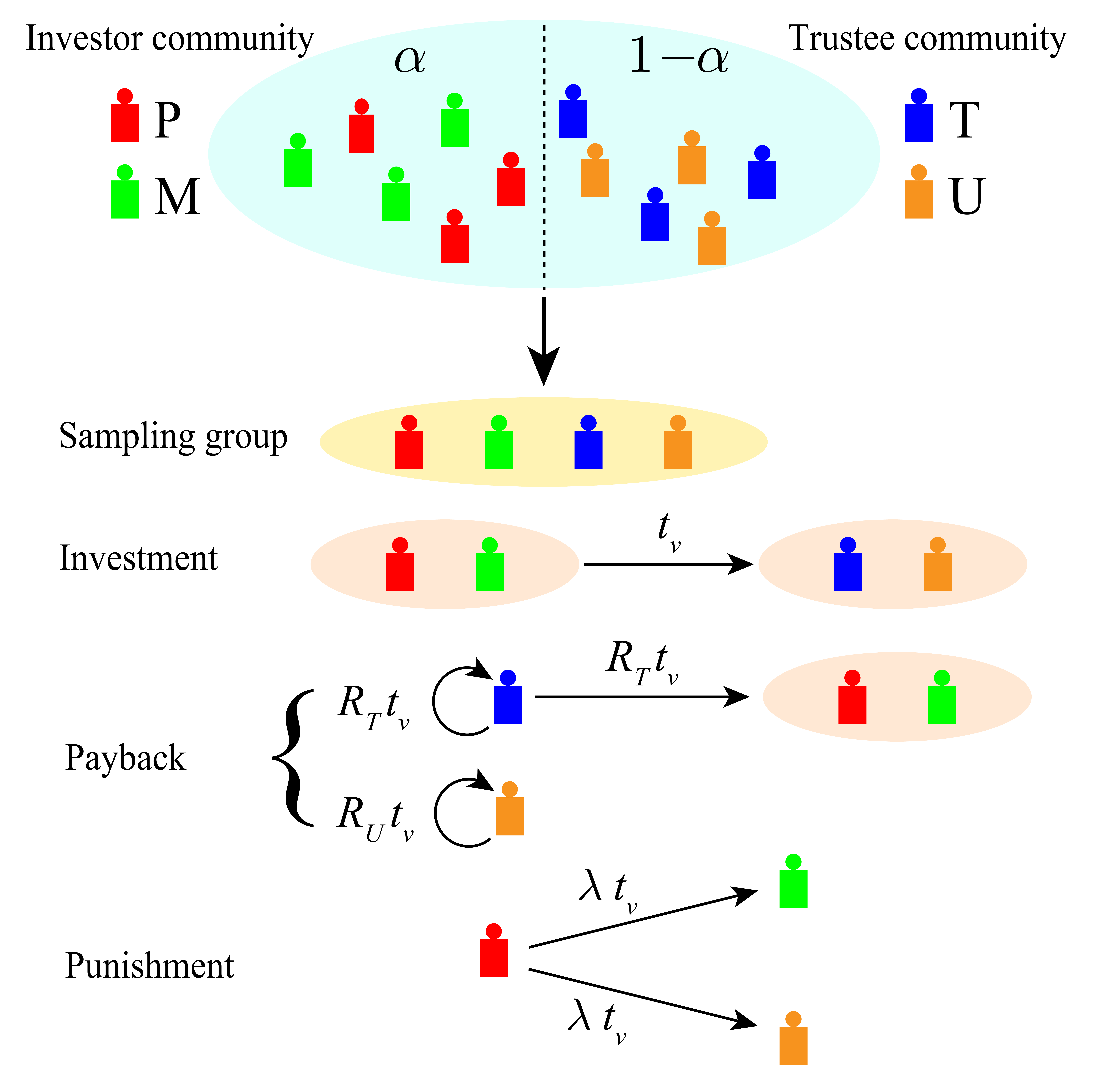}
	\caption{Illustration of the $N$-player trust game in a hierarchical population with punishing investors. First, $N$ individuals are randomly sampled to form a group for playing the game. Second, the investors pay an amount of investment to the trustees. Then, each trustee decides whether to return the received funds. Finally, each punishing investor punishes normal investors and untrustworthy trustees.}
	\label{fig1}
\end{figure}

We denote the number of strategy $j$ in the group by $k_j$ ($j$ = $P$, $M$, $T$, or $U$). According to the standard NTG protocol~\cite{chica2019effects,fang2021evolutionary}, an investor pays $t_v$ to the trustees, which means that the total fund is $(k_P+k_M)t_v$. Each trustee then receives an equal division of this total fund as $(k_P+k_M)t_v/(k_T+k_U)$. After that, a trustworthy trustee returns the received fund multiplied by $R_T$ to the investors, and meanwhile keeps the same amount $R_T(k_P+k_M)t_v/(k_T+k_U)$. However, an untrustworthy trustee returns nothing and finally has the received fund multiplied by $R_U$, which is then $R_U(k_P+k_M)t_v/(k_T+k_U)$, where $1<R_T<R_U<2R_T$. Here, we define the temptation to defect ratio as $r=(R_U-R_T)/R_T \in (0,1)$. We would like to point out that the main difference between the two types of investors is that a punishing investor will curb the trustee's untrustworthy behavior via an additional effort; while a normal investor will not participate in the punishment, but will enjoy its positive consequence. Therefore strategy $M$ could be considered as a sort of second-order free-rider, whose behavior is also sanctioned by player $P$. For simplicity, we consider symmetrical punishment, which means that a punishing investor takes out a budget $\lambda t_v$ to incur a punishment $\lambda t_v/k_U$ on each of the untrustworthy trustees in the group (if any), and also takes out the same budget $\lambda t_v$ to incur a punishment $\lambda t_v/k_M$ on each of the normal investors in the group (if any), where $\lambda >0$ represents the punishment intensity. In order to help readers intuitively understand the interaction process in our game model, we present an illustration figure as shown in Fig.~\ref{fig1}.

Accordingly, the payoffs of punishing investors, normal investors, trustworthy trustees, and untrustworthy trustees from the interacting group can be respectively written as
\begin{equation}
	\pi_{P}=\left\{
	\begin{array}{rcl}
		0,& & \text{if} \ {N_P+N_M=N-1} \text{,}\ \\
		\frac{R_TN_T}{N-1-N_P-N_M}t_v-t_v, & & \text{if} \ {N_U=0 \ \text{,} \ N_M=0 \ \text{and} \ N_P \neq N-1} \text{,}\ \\
		\frac{R_TN_T}{N-1-N_P-N_M}t_v-(1+ \lambda)t_v, & & \text{if} \ {N_U=0 \ \text{,} \ N_M \neq 0 \ \text{and} \ N_P+N_M \neq N-1} \text{,}\ \\
		\frac{R_TN_T}{N-1-N_P-N_M}t_v-(1+ \lambda)t_v, & & \text{if} \ {N_U \neq 0 \ \text{,} \ N_M=0 \ \text{and} \ N_P \neq N-1} \text{,}\ \\
		\frac{R_TN_T}{N-1-N_P-N_M}t_v-(1+2 \lambda)t_v, & & \text{otherwise,}
	\end{array} \right.
\end{equation}
	
\begin{equation}
	\pi_{M}=\left\{
	\begin{array}{rcl}
		0,& & \text{if} \ {N_P+N_M=N-1} \text{,}\ \\
		\frac{R_TN_T}{N-1-N_P-N_M}t_v-t_v-\frac{\lambda N_P  }{N_M+1}t_v, & & \text{otherwise,}
	\end{array} \right.
\end{equation}
	
\begin{equation}
	\pi_{T}=
	\begin{array}{rcl}
		\frac{R_T(N_P+N_M)}{N-N_P-N_M}t_v,
	\end{array}
\end{equation}
and
\begin{equation}
	\pi_{U}=
	\begin{array}{rcl}
		\frac{R_U(N_P+N_M)}{N-N_P-N_M}t_v-\frac{\lambda N_P }{N-N_P-N_M-N_T}t_v,
	\end{array}
\end{equation}
where $N_P$, $N_M$, $N_T$, and $N_U$ represent the number of punishing investors, normal investors, trustworthy trustees, and untrustworthy trustees among another $N-1$ players in the group, respectively.

The evolutionary behavior in the population playing the trust game could be studied by replication dynamics \cite{taylor1978evolutionary}. We denote the fraction of punishing investors, normal investors, trustworthy trustees, and untrustworthy trustees in the population by $x_i$, $y_i$, $x_t$, and $y_t$, respectively. Accordingly we have  $x_i+y_i=\alpha$ and $x_t+y_t=1-\alpha$. Since the population is hierarchical, investors or trustees can only learn from each other, hence there is no transition between the investor community and the trustee community. That is, strategy imitation only alters the fraction of punishing investors and normal investors in the investor community, as well as the fraction of trustworthy trustees and untrustworthy trustees in the trustee community, but cannot change the total fractions of investors or trustees in the population.
The principal goal of present study is to reveal how the efficiency of punishment may depend on the mentioned fractions. Accordingly, the evolution of trust can be described by the following equations \cite{huang2018evolution, liu2018evolution}
\begin{equation}
	\left\{
	\begin{array}{l}
		\dot{x_i}=x_i(f_P-\phi_i),\\
		\dot{y_i}=y_i(f_M-\phi_i),\\
		\dot{x_t}=x_t(f_T-\phi_t),\\
		\dot{y_t}=y_t(f_U-\phi_t),
	\end{array} \right.
\end{equation}
where $f_j$ denotes the expected payoff of strategy $j$ ($j$ = $P$, $M$, $T$, or $U$), $\phi_i=(x_if_P+y_if_M)/\alpha$ and $\phi_t=(x_tf_T+y_tf_U)/(1-\alpha)$ represent the average payoff of investors and trustees, respectively.

In particular, the expected payoff of strategy $P$ can be given by
\begin{equation}
	\begin{aligned}
		f_{P} &= \sum_{N_{P} = 0}^{N-1}\sum_{N_{M} = 0}^{N-N_{P}-1}\sum_{N_{T} =0}^{N-N_P-N_M-1}\binom{N-1}{N_{P}}\binom{N-N_{P}-1}{N_{M}}\binom{N-N_{P}-N_{M}-1}{N_{T}}\\
		&\times x_i^{N_{P}}(\alpha-x_i)^{N_{M}}x_t^{N_{T}}(1-\alpha-x_t)^{N-N_P-N_M-N_T-1}\pi_{P}\\
		&= x_tR_T\frac{1-\alpha ^{N-1}}{1-\alpha}t_v-(1+2\lambda)(1-\alpha^{N-1})t_v\\
		&+\lambda((1-\alpha +x_i)^{N-1}-x_i^{N-1}+(\alpha +x_t)^{N-1}-\alpha^{N-1}) t_v,
	\end{aligned}
\end{equation}
while the expected payoff of strategy $M$ is given by
\begin{equation}
	\begin{aligned}
		f_{M} &= \sum_{N_{P} = 0}^{N-1}\sum_{N_{M} = 0}^{N-N_{P}-1}\sum_{N_{T} =0}^{N-N_P-N_M-1}\binom{N-1}{N_{P}}\binom{N-N_{P}-1}{N_{M}}\binom{N-N_{P}-N_{M}-1}{N_{T}}\\
		&\times x_i^{N_{P}}(\alpha-x_i)^{N_{M}}x_t^{N_{T}}(1-\alpha-x_t)^{N-N_P-N_M-N_T-1}\pi_{M}\\
		&=x_tR_T\frac{1-\alpha ^{N-1}}{1-\alpha}t_v-(1-\alpha^{N-1})t_v\\
		&-\lambda x_i\frac{1-(1-\alpha +x_i)^{N-1}}{\alpha -x_i}t_v+\lambda x_i\frac{\alpha^{N-1}-x_i^{N-1}}{\alpha-x_i}t_v\,.
	\end{aligned}
\end{equation}
The expected payoff of strategy $T$ can be given by
\begin{equation}
	\begin{aligned}
		f_{T} &= \sum_{N_{P} = 0}^{N-1}\sum_{N_{M} = 0}^{N-N_{P}-1}\sum_{N_{T} =0}^{N-N_P-N_M-1}\binom{N-1}{N_{P}}\binom{N-N_{P}-1}{N_{M}}\binom{N-N_{P}-N_{M}-1}{N_{T}}\\
		&\times x_i^{N_{P}}(\alpha-x_i)^{N_{M}}x_t^{N_{T}}(1-\alpha-x_t)^{N-N_P-N_M-N_T-1}\pi_{T}\\
		&=\alpha R_T\frac{1-\alpha^{N-1}}{1-\alpha}t_v,
	\end{aligned}
\end{equation}
while the expected payoff of strategy $U$ is given by
\begin{equation}
	\begin{aligned}
		f_{U} &= \sum_{N_{P} = 0}^{N-1}\sum_{N_{M} = 0}^{N-N_{P}-1}\sum_{N_{T} =0}^{N-N_P-N_M-1}\binom{N-1}{N_{P}}\binom{N-N_{P}-1}{N_{M}}\binom{N-N_{P}-N_{M}-1}{N_{T}}\\
		&\times x_i^{N_{P}}(\alpha-x_i)^{N_{M}}x_t^{N_{T}}(1-\alpha-x_t)^{N-N_P-N_M-N_T-1}\pi_{U}\\
		&=\alpha (r+1)R_T\frac{1-\alpha^{N-1}}{1-\alpha}t_v-\lambda x_i\frac{1-(\alpha+x_t)^{N-1}}{1-\alpha-x_t}t_v.
	\end{aligned}
\end{equation}

In the following we study the evolutionary dynamics of trust by analyzing the existence and stabilities of equilibria in the defined model, and show that there exists stable coexistence state between punishing investors and trustworthy trustees. Furthermore, we investigate how the key parameters influence the attraction domain of the coexistence state.

\section{Results}
\subsection{Analysis of stable equilibria}
As we have $x_i+y_i=\alpha$ and $x_t+y_t=1-\alpha$ constraints, the replicator equations can be simplified as
\begin{equation}
	\left\{
	\begin{array}{rcl}
		\dot{x_i}=x_i(f_P-\phi_i),\\
		\dot{x_t}=x_t(f_T-\phi_t).
	\end{array} \right.
\end{equation}

Since the mathematical expression of the  payoffs of strategies in the population is highly non-linear, it is difficult to prove whether there is an interior equilibrium. However, if it exists, we can then prove that it is unstable (see Appendix~C for details).
\begin{figure}[h!]
	\centering
	\includegraphics[width=14cm]{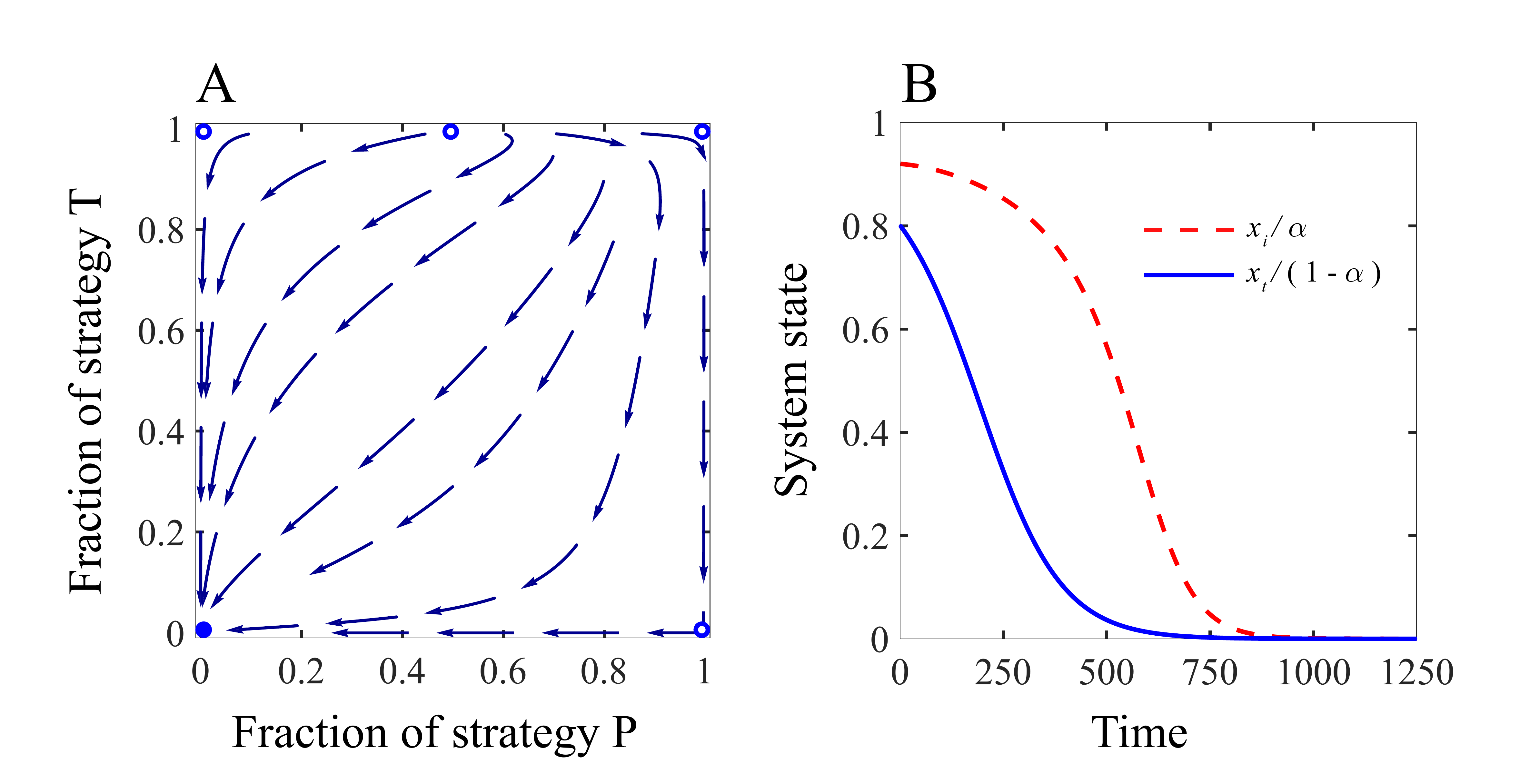}
	\caption{Panel~A: Phase flows of the system for $\alpha < \alpha^*$ and $ \lambda<rR_T(1-\alpha^{N-1})/((N-1)(1-\alpha))$. Arrows indicate the direction of evolution. The stable fixed point in the left-down corner is depicted by a solid blue circle, while unstable fixed points are marked by open blue circles. Panel~B: Time evolution of the fraction of strategy $P$ (red-dashed line) and strategy $T$ (blue-solid line). The system in Panel~B finally evolves into the stable equilibrium of $M+U$ strategies. Parameters: $\alpha= 0.1$, $\lambda= 0.01$, $r=0.05$, $R_T=2$, $N=10$, and $t_v=1$.}
	\label{fig2}
\end{figure}

Because of the mentioned difficulties, we mainly focus on the existence and stabilities of boundary equilibria. As shown in Appendix~A, there are at most seven boundary equilibria: equilibrium $M+U$ (i.e., fixed point $(0,0)$); equilibrium $M+T$ (i.e., fixed point $(0,1-\alpha)$); equilibrium $P+U$ (i.e., fixed point $(\alpha,0)$); equilibrium $P+T$ (i.e., fixed point $(\alpha,1-\alpha)$); equilibrium $P+T+U$ (i.e., fixed point $(\alpha,\overline{x}_{t1})$), where $\overline{x}_{t1}$ is the unique solution of $\lambda \frac{1-(\alpha+x)^{N-1}}{1-\alpha-x}-rR_T\frac{1-\alpha^{N-1}}{1-\alpha}=0$; equilibrium $P+M+U$ (i.e., fixed point $(\overline{x}_{i1},0)$), where $\overline{x}_{i1}$ is the unique solution of $x\frac{1-(1-\alpha+x)^{N-1}}{\alpha-x}-x\frac{\alpha^{N-1}-x^{N-1}}{\alpha-x}+(1-\alpha+x)^{N-1}-x^{N-1}-2+2\alpha^{N-1}=0$; and equilibrium $P+M+T$ (i.e., fixed point $(\overline{x}_{i2},1-\alpha)$), where $\overline{x}_{i2}$ is the unique solution of $x\frac{1-(1-\alpha+x)^{N-1}}{\alpha-x}-x\frac{\alpha^{N-1}-x^{N-1}}{\alpha-x}+(1-\alpha+x)^{N-1}-x^{N-1}-1+\alpha^{N-1}=0$.

Next, we investigate the existence and stabilities of boundary equilibria in different parameter regions. For simplicity, we denote the unique solution of $(N-1)\alpha-(N-2)\alpha^{N-1}=1$ by $\alpha^*$ (in particular, $\alpha^*=1$ for $N=3$). Accordingly, we have the following conclusions.

Case~$(1)$: If $\alpha < \alpha^*$ and $ \lambda<rR_T(1-\alpha^{N-1})/((N-1)(1-\alpha))$, there are five boundary equilibria: $M+U$, $M+T$, $P+U$, $P+T$, and $P+M+T$ (i.e., fixed points $(0,0)$, $(0,1-\alpha)$, $(\alpha,0)$,  $(\alpha,1-\alpha)$, and $(\overline{x}_{i2},1-\alpha)$). $M+U$ (i.e., fixed point $(0,0)$) is the unique stable equilibrium (see Appendix~B.1 for details).

Furthermore, we provide some numerical calculations for the above theoretical results. As shown in Fig.~\ref{fig2}, there are five boundary fixed points, while only $(0,0)$ is stable. In this case, for each trustee, the maximum expected loss $\lambda \alpha(N-1)t_v$ for defecting is always less than the expected gain $\alpha rR_Tt_v(1-\alpha^{N-1})/(1-\alpha)$ for doing so. Thus, the trustees always choose to defect and strategy $T$ dies out. Furthermore, due to the small fraction of investors, punishing investors cannot dominate over normal investors when there are only a few trustworthy trustees. Accordingly, independently of the initial conditions the population will always evolve into the equilibrium state where normal investors and untrustworthy trustees coexist.

Case~$(2)$: If $\alpha > \alpha^*$ and $ \lambda<rR_T(1-\alpha^{N-1})/((N-1)(1-\alpha))$, there are six boundary equilibria: $M+U$, $M+T$, $P+U$, $P+T$, $P+M+U$, and $P+M+T$ (i.e., fixed points $(0,0)$, $(0,1-\alpha)$, $(\alpha,0)$, $(\alpha,1-\alpha)$, $(\overline{x}_{i1},0)$, and $(\overline{x}_{i2},1-\alpha)$). $M+U$ and $P+U$ (i.e., fixed points $(0,0)$ and $(\alpha,0)$) are both stable equilibria, while other fixed points are unstable (see Appendix~B.2 for details).

As shown in Fig.~\ref{fig3}, our numerical calculations confirm that there are six boundary fixed points, while $(0,0)$ and $(\alpha,0)$ are stable. Similar to the case~$(1)$, since $ \lambda<rR_T(1-\alpha^{N-1})/((N-1)(1-\alpha))$, the trustees always return nothing. But the evolutionary outcome, whether punishing investors can dominate over normal investors or not, depends on the initial states. Accordingly, the population can terminate into $M+U$ or $P+U$ equilibrium depending on the initial conditions. This effect is illustrated in Panel~B and Panel~C of Fig.~\ref{fig3}.
\begin{figure}[h!]
	\centering
	\includegraphics[width=16cm]{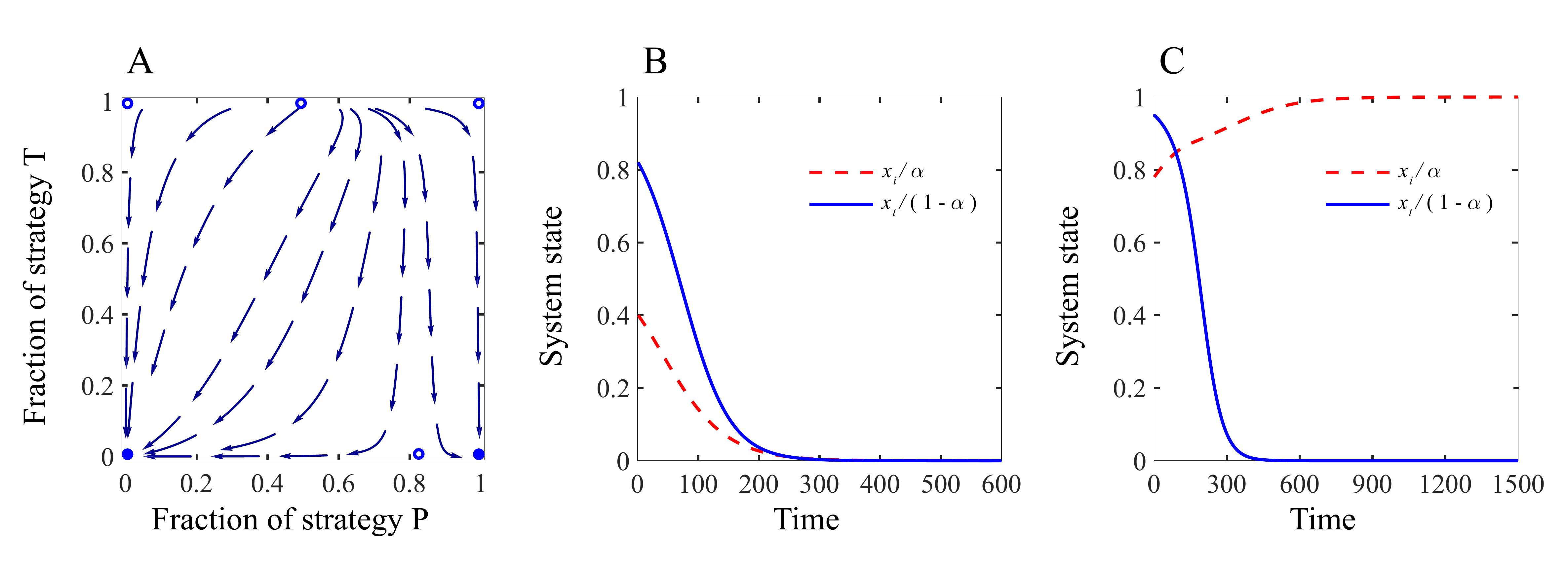}
	\caption{Panel~A: Phase flows of the system for $\alpha > \alpha^*$ and $ \lambda<rR_T(1-\alpha^{N-1})/((N-1)(1-\alpha))$. Arrows indicate the direction of evolution. Stable fixed points in the left-down and right-down corners are depicted by solid blue circles, while unstable fixed points are marked by open blue circles. The importance of initial conditions is demonstrated in Panel~B and Panel~C where we plot the time evolution of the fraction of strategy $P$ (red-dashed lines) and strategy $T$ (blue-solid lines). The system in Panel~B finally evolves into the stable equilibrium $M+U$ state, while the system in Panel~C finally evolves into the stable equilibrium $P+U$. Parameters:  $\alpha= 0.2$, $\lambda=0.01$, $r=0.05$, $R_T=2$, $N=10$, and $t_v=1$.}
	\label{fig3}
\end{figure}

Case~$(3)$: If $\alpha < \alpha^*$ and  $rR_T(1-\alpha^{N-1})/((N-1)(1-\alpha))<\lambda<rR_T$, there are six boundary equilibria: $M+U$, $M+T$, $P+U$, $P+T$, $P+T+U$, and $P+M+T$ (i.e., fixed points $(0,0)$, $(0,1-\alpha)$, $(\alpha,0)$, $(\alpha,1-\alpha)$, $(\alpha,\overline{x}_{t1})$, and $(\overline{x}_{i2},1-\alpha)$). $M+U$ and $P+T$ (i.e., fixed points $(0,0)$ and $(\alpha,1-\alpha)$) are both stable equilibria, other fixed points are unstable (see Appendix~B.3 for details).
\begin{figure}[h!]
	\centering
	\includegraphics[width=16cm]{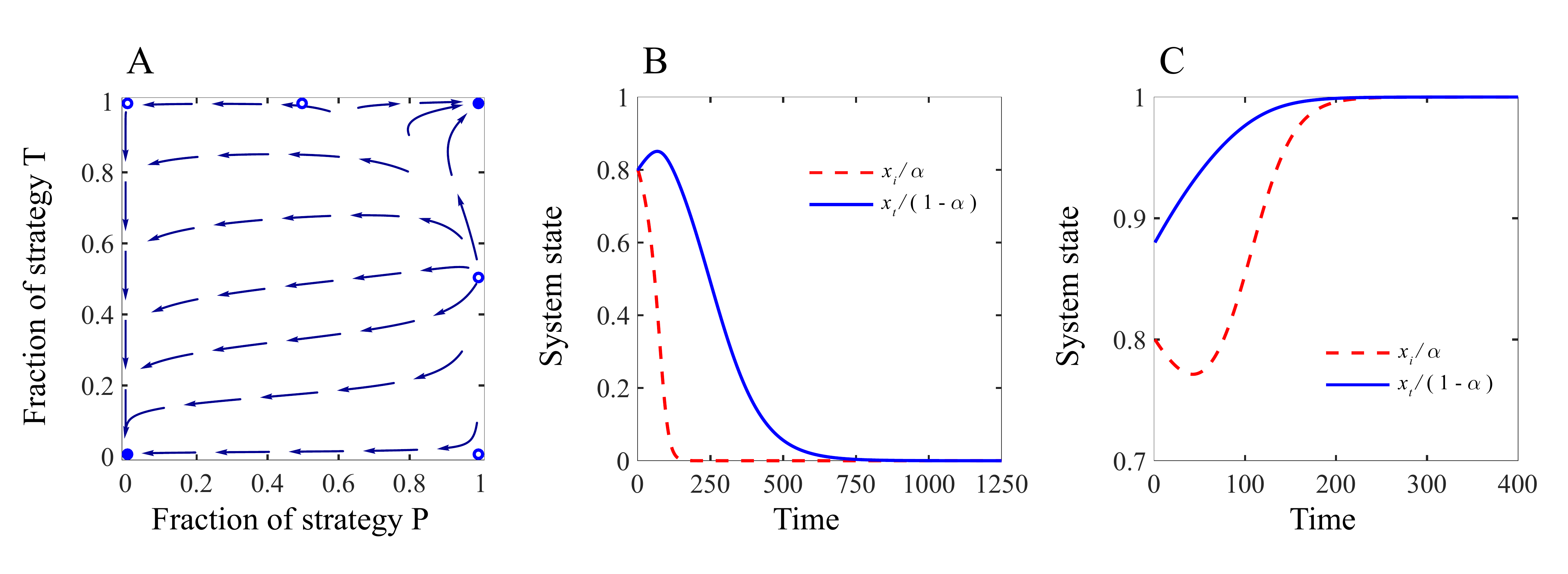}
	\caption{Panel~A: Phase flows of the system for $\alpha < \alpha^*$ and  $rR_T(1-\alpha^{N-1})/((N-1)(1-\alpha))<\lambda<rR_T$. Arrows indicate the direction of evolution. Stable fixed points in the left-down and right-up corners are depicted by solid blue circles, while unstable fixed points are marked by open blue circles. The importance of initial conditions is illustrated in Panel~B and Panel~C where time evolution of the fraction of strategy $P$ (red-dashed lines) and strategy $T$ (blue-solid lines) is plotted. The system in Panel~B finally evolves into the stable equilibrium $M+U$, while the system in Panel~C finally evolves into the stable equilibrium $P+T$. Parameters: $\alpha= 0.1$, $\lambda=0.05$, $r=0.05$, $R_T=2$, $N=10$, and $t_v=1$.}
	\label{fig4}
\end{figure}

Numerical calculations are presented in Fig.~\ref{fig4}. We can see that there are six boundary fixed points, while $(0,0)$ and $(\alpha,1-\alpha)$ are stable. Compared with the case $(1)$, the punishment to the untrustworthy trustees is increased. Accordingly, the population may finally reach the stable coexistence state of punishing investors and trustworthy trustees. Similar to the previous case, the initial conditions have special importance, hence the population can evolve either into the equilibrium $M+U$ or $P+T$ solution.

Case~$(4)$: If $\alpha > \alpha^*$ and $rR_T(1-\alpha^{N-1})/((N-1)(1-\alpha))<\lambda<rR_T$, there are seven boundary equilibria: $M+U$, $M+T$, $P+U$, $P+T$, $P+T+U$, $P+M+U$, and $P+M+T$ (i.e.,  fixed points $(0,0)$, $(0,1-\alpha)$, $(\alpha,0)$, $(\alpha,1-\alpha)$, $(\alpha,\overline{x}_{t1})$, $(\overline{x}_{i1},0)$, and $(\overline{x}_{i2},1-\alpha)$). While $M+U$, $P+U$, and $P+T$ (i.e., fixed points $(0,0)$, $(\alpha,0)$, and $(\alpha,1-\alpha)$) are stable equilibria, the remaining four are unstable (see Appendix~B.4 for details).
\begin{figure}[h!]
	\centering
	\includegraphics[width=17cm]{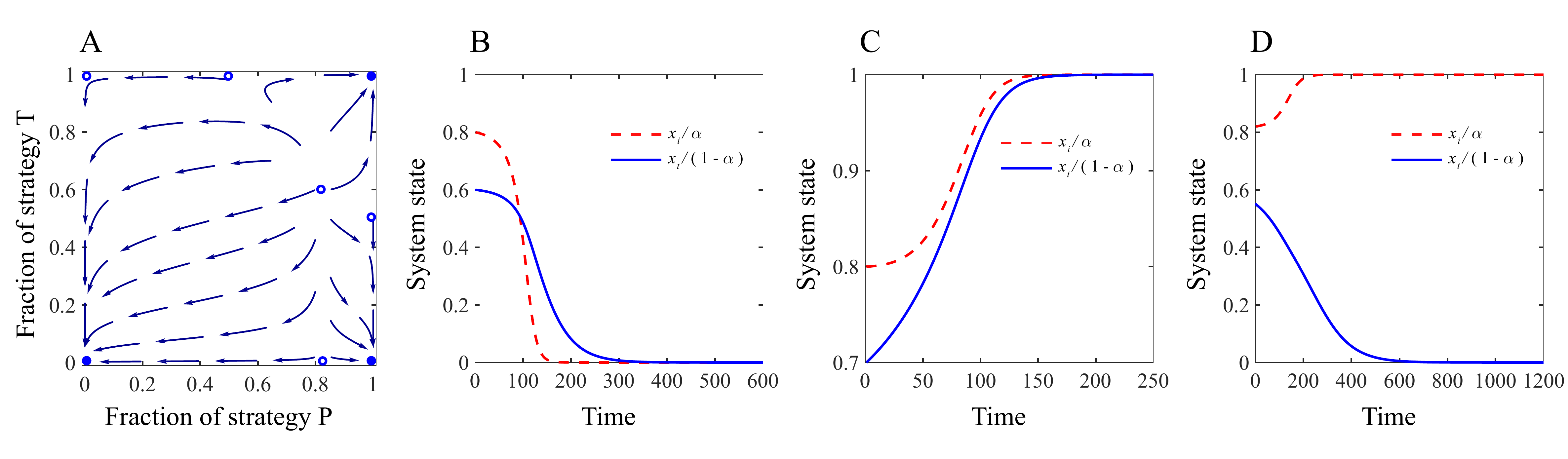}
	\caption{Panel~A: Phase flows of the system for $\alpha > \alpha^*$ and $rR_T(1-\alpha^{N-1})/((N-1)(1-\alpha))<\lambda<rR_T$. Arrows indicate the direction of evolution. Stable fixed points are depicted by solid blue circles, while unstable fixed points are marked by open blue circles. The final destination depends sensitively on the initial conditions, which is illustrated in Panels~B, C, and D. Here we plot the time evolution of the fraction of strategy $P$ (red-dashed lines) and strategy $T$ (blue-solid lines). In the first case the system evolves into the stable equilibrium $M+U$, while Panel~C shows that the system evolves into the stable equilibrium $P+T$. In Panel~D the system evolves into the stable equilibrium $P+U$ state.  Parameters:  $\alpha= 0.2$, $\lambda=0.05$, $r=0.05$, $R_T=2$, $N=10$, and $t_v=1$.}
	\label{fig5}
\end{figure}

As shown in Fig.~\ref{fig5}, there are seven boundary fixed points and one interior fixed point, while boundary fixed points $(0,0)$, $(\alpha,0)$, and $(\alpha,1-\alpha)$ are stable. Compared with the case~$(3)$, the fraction of investors is increased. Thus, punishing investors may outcompete normal investors, even if trustworthy trustees are few. Accordingly, depending on the initial conditions the population can evolve into one of the equilibrium $M+U$, $P+U$, or $P+T$ points.

Case~$(5)$: If $\alpha \leq \alpha^*$ and $\lambda>rR_T$, there are five boundary equilibria: $M+U$, $M+T$, $P+U$, $P+T$, and $P+M+T$ (i.e., fixed points $(0,0)$, $(0,1-\alpha)$, $(\alpha,0)$, $(\alpha,1-\alpha)$, and $(\overline{x}_{i2},1-\alpha)$). Here $M+U$ and $P+T$ (i.e., fixed points $(0,0)$ and $(\alpha,1-\alpha)$) are both stable equilibria, others are unstable (see Appendix~B.5 for details).

\begin{figure}[h!]
	\centering
	\includegraphics[width=16cm]{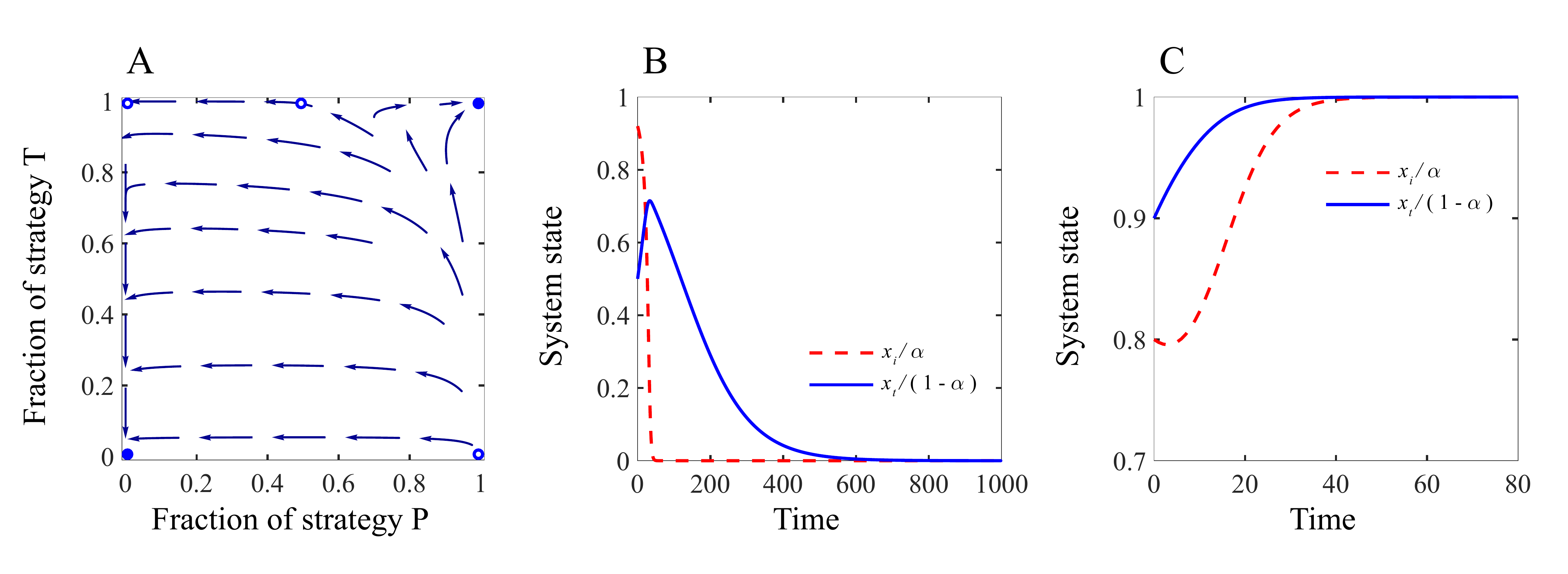}
	\caption{Panel~A: Phase flows of the system for $\alpha \leq \alpha^*$ and $\lambda>rR_T$. Arrows indicate the direction of evolution. Stable fixed points in the opposite corners are depicted by solid blue circles, while unstable fixed points are marked by open blue circles. Panels~B and~C show the time evolution of the fraction of strategy $P$ (red-dashed lines) and strategy $T$ (blue-solid lines). The final destination is either $M+U$ or $P+T$. Parameters:  $\alpha= 0.1$, $\lambda=0.2$, $r=0.05$, $R_T=2$, $N=10$, and $t_v=1$.}
	\label{fig6}
\end{figure}

The fixed points are presented in Fig.~\ref{fig6}. Compared to the case~$(4)$, the punishment intensity is higher, so even if the fraction of investors is small, there does not exist stable coexistence between punishing investors and untrustworthy trustees. Accordingly, in dependence on the initial conditions, the population will evolve into the equilibrium $M+U$ or $P+T$.

Case~$(6)$: If $\alpha > \alpha^*$ and $\lambda>rR_T$, there are six boundary equilibria: $M+U$, $M+T$, $P+U$, $P+T$, $P+M+U$, and $P+M+T$ (i.e., fixed points $(0,0)$, $(0,1-\alpha)$, $(\alpha,0)$, $(\alpha,1-\alpha)$, $(\overline{x}_{i1},0)$, and $(\overline{x}_{i2},1-\alpha)$). $M+U$ and $P+T$ (i.e., fixed points $(0,0)$ and $(\alpha,1-\alpha)$) are both stable equilibria, while the remaining four are unstable (see Appendix~B.6 for details).

\begin{figure}[h!]
	\centering
	\includegraphics[width=16cm]{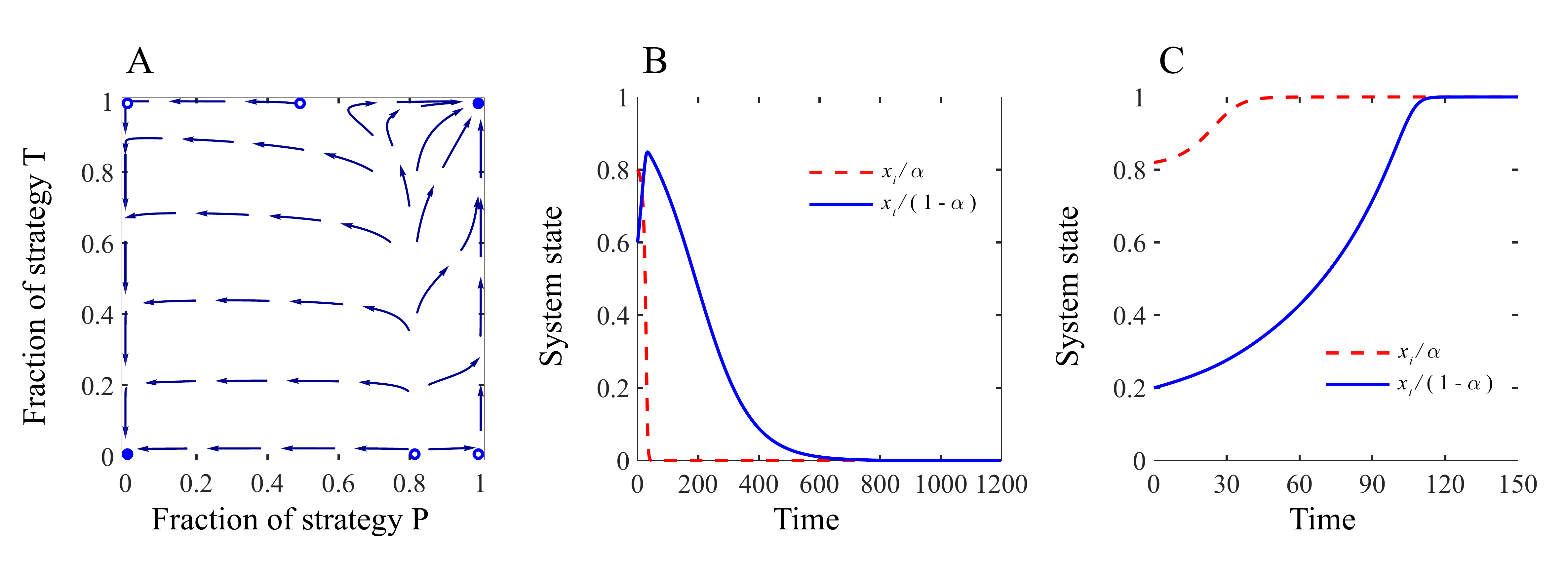}
	\caption{Panel~A: Phase flows of the system for $\alpha > \alpha^*$ and $\lambda>rR_T$. Arrows indicate the direction of evolution. Stable fixed points in the opposite corners are depicted by solid blue circles, while unstable fixed points are marked by open blue circles. Panel~B and Panel~C: Time evolution of the fraction of strategy $P$ (red-dashed lines) and strategy $T$ (blue-solid lines). In the first case the system evolves into the stable $M+U$ solution, while Panel~C illustrates the destination into the stable $P+T$ equilibrium point. Parameters: $\alpha= 0.1$, $\lambda=0.2$, $r=0.05$, $R_T=2$, $N=20$, and $t_v=1$.}
	\label{fig7}
\end{figure}

Numerical calculations can be found in Fig.~\ref{fig7} where we present the time evolution to the stable $(0,0)$ and $(\alpha,1-\alpha)$ fixed points. Similar to the case~$(5)$, the coexistence between punishing investors and untrustworthy trustees is unstable due to the high punishment intensity. Accordingly, in dependence on the initial conditions, the population will evolve either into the equilibrium $M+U$ or $P+T$ solution.

\begin{figure}[h!]
	\centering
	\includegraphics[width=12cm]{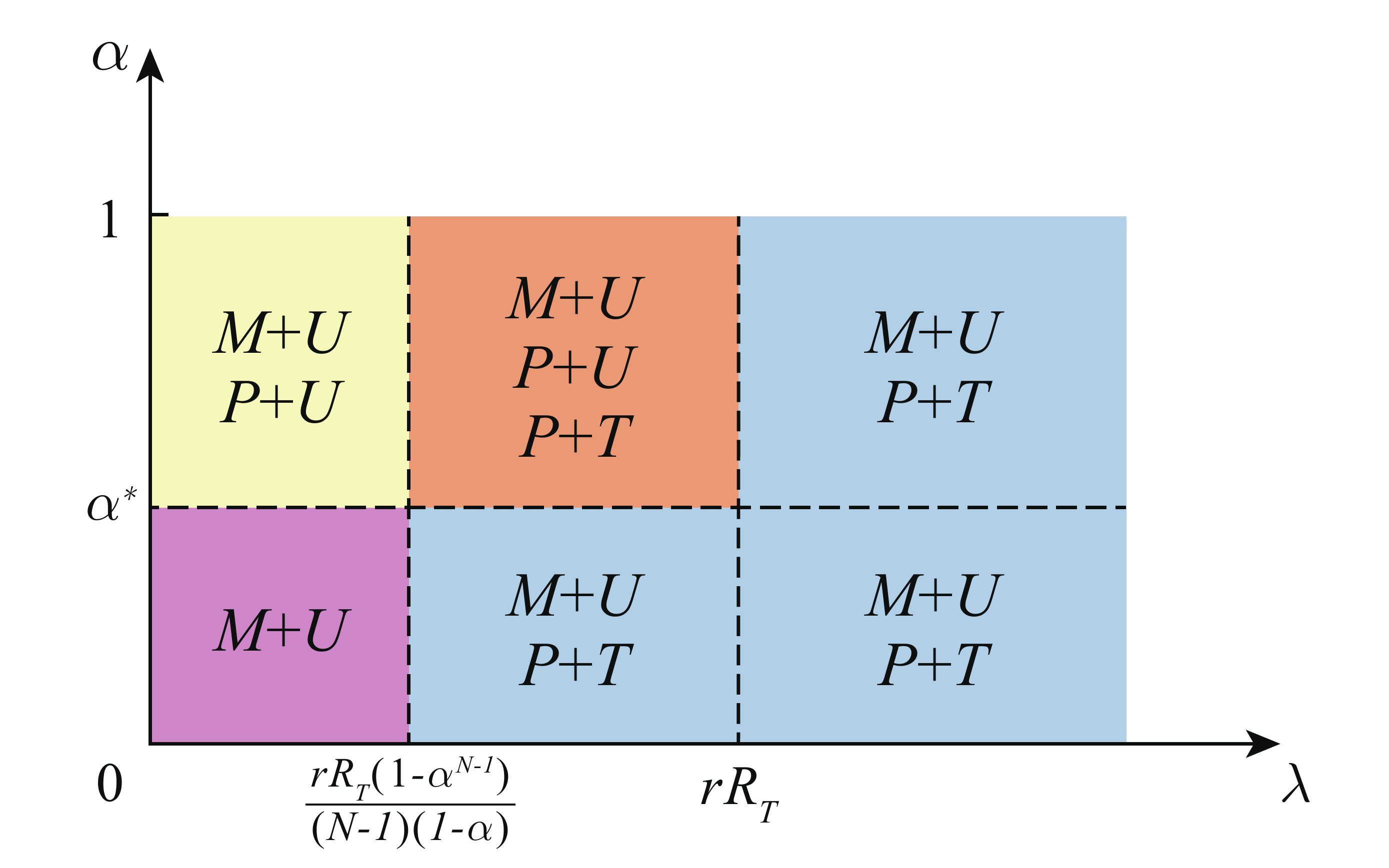}
	\caption{A representative plot of evolutionary outcomes on the phase plane. We use different colors to distinguish different possible evolutionary outcomes in the $\lambda-\alpha$ parameter space and show all evolutionary stable equilibria, where $\alpha^*$ is the unique solution of $(N-1)\alpha-(N-2)\alpha^{N-1}=1$ (in particular, $\alpha^*=1$ for $N=3$).}
	\label{fig8}
\end{figure}

To summarize our findings and give a comprehensive survey to our readers about the possible solutions, we present an illustrative plot of the dynamical regimes in the $\lambda-\alpha$ parameter space and use different colors to distinguish possible evolutionary destinations in Fig.~\ref{fig8}.

\subsection{Impacts of key parameters on the attraction domain of the coexistence state}

\begin{figure}[h!]
	\centering
	\includegraphics[width=15cm]{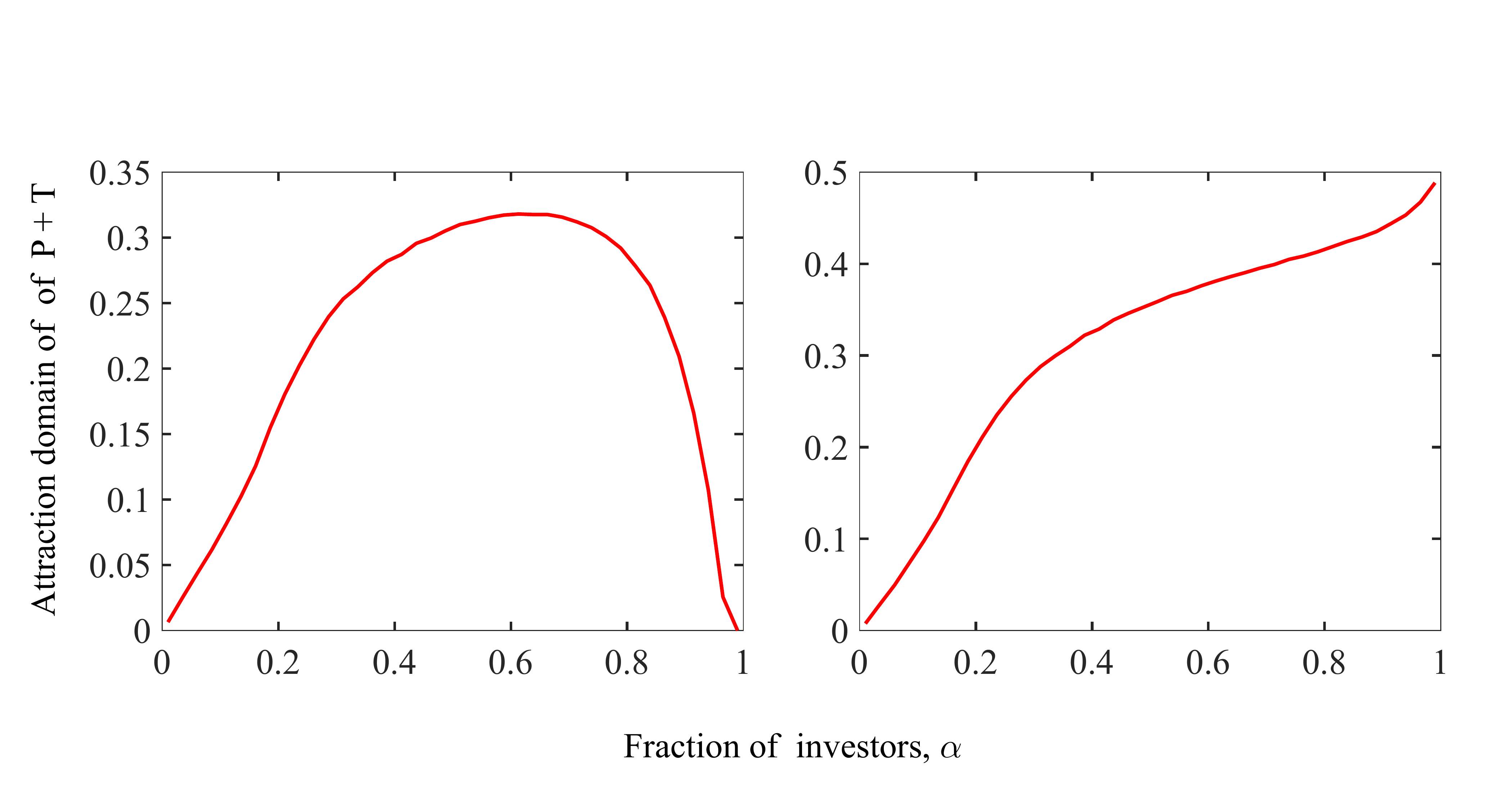}
	\caption{The attraction domain of the coexistence state of punishing investors and trustworthy trustees as a function of the $\alpha$ fraction of investors. Panel~A shows the low punishment region where $\lambda=0.14$, while Panel~B is for high punishment $\lambda=0.5$ case. Other parameters are $r=0.05$, $R_T=3$, $N=7$, and $t_v=1$ for both panels.}
	\label{fig9}
\end{figure}

\begin{figure}[h!]
	\centering
	\includegraphics[width=12cm]{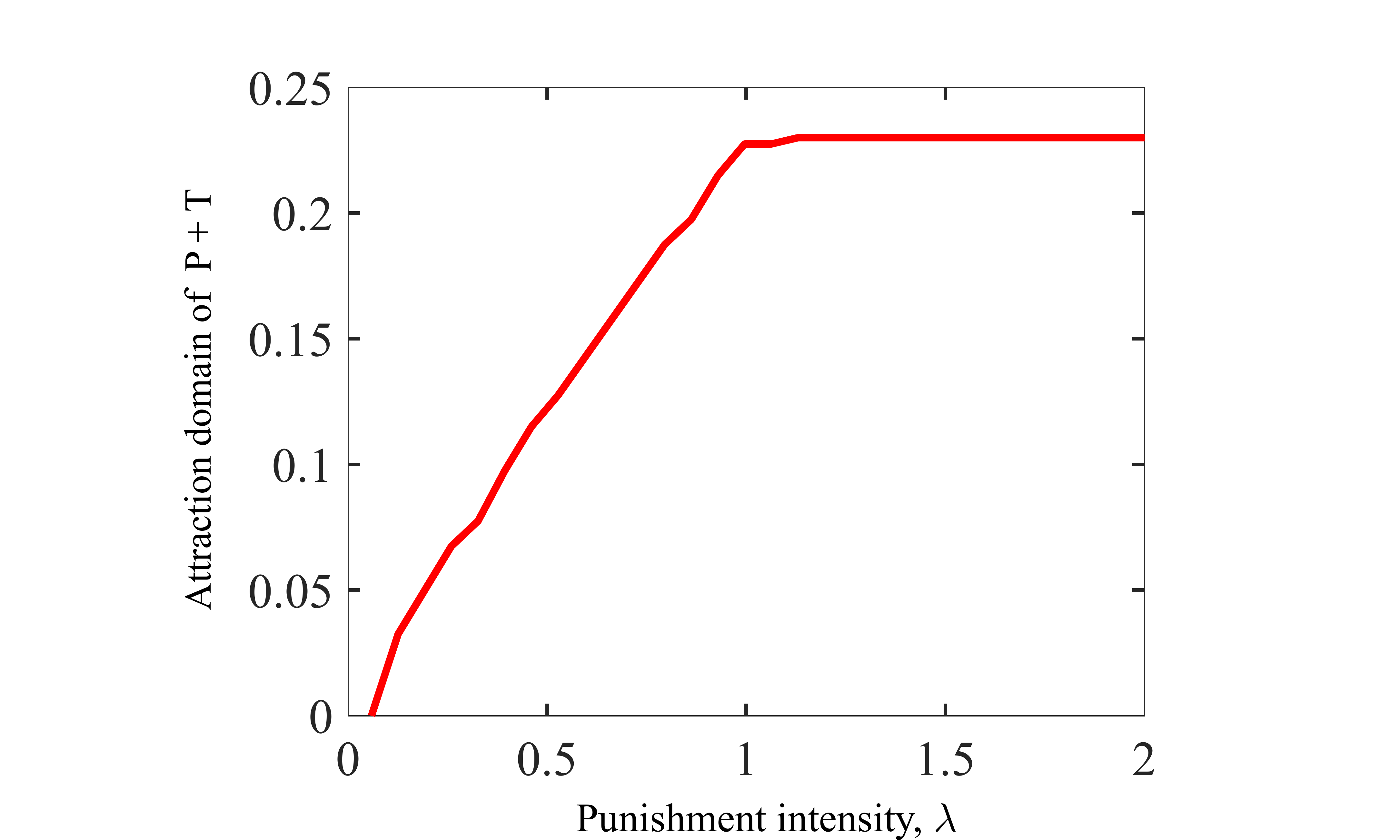}
	\caption{The attraction domain of the coexistence state of punishing investors and trustworthy trustees as a function of the punishment intensity $\lambda$. It suggests that it is useful to increase the level of punishment, but just until a certain point. Parameters: $\alpha=0.1$, $r=0.5$, $R_T=2$, $N=20$, and $t_v=1$.}
	\label{fig10}
\end{figure}

According to the analysis presented above, we have shown that there exists stable coexistence state of punishing investors and trustworthy trustees, hence trust can emerge due to the punishment. In the following we explore how key parameters affect the attraction domain of the coexistence state. In particular, we focus on parameter $\alpha$ which characterizes the fraction of investors in our hierarchical population and parameter $\lambda$. The latter determines the punishment intensity. In other words, their combination represents the applied incentive to improve the trust level.

We begin our discussion by presenting the attraction domain of the coexistence state in dependence of the $\alpha$ fraction of investors. For low punishment intensity, as illustrated in Fig.~\ref{fig9}A, the $\alpha$-dependence is not monotonous. First the attraction domain grows monotonously as we increase the portion of investors until reaching a maximum value, and then decreases. This behavior highlights that an intermediate fraction of investors is more beneficial to the evolution of trust when the punishment intensity is low. In the opposite case, shown in Fig.~\ref{fig9}B, when the punishment is strong, we can observe a clear correlation between the attraction domain and parameter $\alpha$. This indicates that a higher fraction of investors is always more beneficial to the evolution of trust in these conditions.

Last, we present how the punishment intensity $\lambda$ influences the evolution of trust at a fixed portion of investors. A representative behavior is shown in Fig.~\ref{fig10} where the attraction domain of the coexistence state first increases almost linearly by increasing $\lambda$, and then saturates and remains constant no matter we imply larger $\lambda$ value. This reflects that, an appropriate increase of the punishment intensity can help trust to evolve, but it becomes useless to increase it above a certain point. We note that this observation fits nicely to previous observations about the careful usage of punishment \cite{helbing10,liu19,jiang13}.

\section{Conclusion}

In this work, we have explored the possible consequences of the asymmetrical nature of the $N$-player trust game. More precisely, we allow for investor and trustee communities to be differently large who can learn from each other within their groups. Beside the usual trustworthy and untrustworthy trustee options, we introduce the choice of investors to punish untrustworthy trustees and those, otherwise normal, investors who deny to implement punishment on untrustworthy individuals. In this way the portion of investor community and the intensity of punishment are the key parameters of our model.

By using replicator equations, we have found that the introduction of punishment can lead to the stable coexistence between punishing investors and trustworthy trustees, which indicates that the evolution of trust can be greatly promoted in this way. Additionally, by means of numerical calculations, we have revealed that for low punishment intensity, an intermediate fraction of investors can help trust evolve better. For strong punishment intensity, however, a higher fraction of investors is always better to reach the desired solution. Last we could also support the observations of related models about conducive role of more severe punishment, which could be useful to a certain point, but not further~\cite{helbing10,liu19,jiang13}. Therefore, we believe that our findings not only reveal the role of punishment in promoting trust in a hierarchical population, but also show how to choose the proper punishment intensity for boosting trust in different cases, which could be helpful to incentive implementation for promoting prosocial behaviors in the realistic world. This aspect could be especially important because applying punishment is always costly, which could be a big obstacle to do it. Therefore, our work may provide some insights into enhancing trust among interacting individuals in a realistic situation of hierarchical populations.

Our present paper mainly investigates the evolutionary dynamics of trust in a hierarchical population from a theoretical perspective. However, it still has some limitations, which could be the scope of further studies. For example, we focused on symmetrical punishment, that is, punishing investors use equal budgets to punish untrustworthy trustees and normal investors. However, given the differences between investors and trustees, the punishments may be asymmetrical. In the light of previous studies about asymmetrical punishment \cite{huang2018evolution,fowler2005altruistic,engel2016symmetric}, a promising extension of our work is thus to investigate the effects of asymmetrical punishment on the evolution of trust for a subsequent work. In addition, in this work we have considered a well-mixed interaction where individuals form interaction groups randomly and temporarily. However, the interactions among individuals are typically not random but rather that they are limited to a set of neighbors and more or less fixed in time. These conditions can be modeled in a structured population, which could be described by a complex interaction network~\cite{kumar2020evolution,hu2021ieeetnse,li2020ieeetii,li2022ieeetkde,li2021acmtkdd,li2020njp,li2021dynamics}.
Notably, the change of these conditions may modify the evolutionary outcomes significantly~\cite{nowak,perc}, hence we may expect new observations in the extended models for future study. Furthermore, since trust plays an important role in realistic human society,
it would be worthwhile to combine the relevant data sets including the experimental data from behavior experiments as well as the empirical data, for verifying the efficiency of our theoretical models where real factors are considered. We believe that future work along these avenues will be valuable and improve our understanding on how trust evolves and emerges in the realistic world.

\section*{Acknowledgments}
This research was supported by the National Natural Science Foundation of China (Grant Nos. 61976048 and 62036002) and the Fundamental Research Funds of the Central Universities of China.

\newpage

\appendix

\section{Analysis of boundary fixed points}
In this section, we investigate the boundary fixed points in more detail.

Since the constraints of $x_i+y_i=\alpha$ and $x_t+y_t=1-\alpha$ are always hold, the replicator equations can be simplified as
\begin{equation}
	\left\{
	\begin{array}{lr}
		\dot{x_i}=x_i(f_P-\phi_i),\\
		\dot{x_t}=x_t(f_T-\phi_t),
	\end{array} \nonumber
	\right.
\end{equation}
where $\phi_i=(x_if_P+y_if_M)/\alpha$ and $\phi_t=(x_tf_T+y_tf_U)/(1-\alpha)$.

Then we have
\begin{equation}
	\left\{
	\begin{array}{lr}
		\dot{x_i}=\frac{1}{\alpha}x_i(\alpha-x_i)(f_P-f_M),\\
		\dot{x_t}=\frac{1}{1-\alpha}x_t(1-\alpha-x_t)(f_T-f_U).
	\end{array}\nonumber
	\right.
\end{equation}

Let $f(x_i,x_t)=x_i\frac{1-(1-\alpha+x_i)^{N-1}}{\alpha-x_i}-x_i\frac{\alpha^{N-1}-x_i^{N-1}}{\alpha-x_i}+(\alpha+x_t)^{N-1}+(1-\alpha+x_i)^{N-1}-x_i^{N-1}-2+\alpha^{N-1}$ and $g(x_i,x_t)=\lambda x_i\frac{1-(\alpha+x_t)^{N-1}}{1-\alpha-x_t}-\alpha rR_T\frac{1-\alpha^{N-1}}{1-\alpha}$, then the replicator equations can be rewritten as
\begin{equation}
	\left\{
	\begin{array}{lr}
		\dot{x_i}=\frac{1}{\alpha}\lambda t_vx_i(\alpha-x_i)f(x_i,x_t),\\
		\dot{x_t}=\frac{1}{1-\alpha}t_vx_t(1-\alpha-x_t)g(x_i,x_t).
	\end{array}\nonumber
	\right.
\end{equation}

It is easy to see that $(0,0)$, $(0,1-\alpha)$, $(\alpha,0)$, and $(\alpha,1-\alpha)$ are boundary fixed points. Below, we explore other boundary fixed points.

Let $x_i=0$, we restrict $x_t(1-\alpha-x_t)\neq 0$. Solving $\dot{x_i}=0$ and $\dot{x_t}=0$ results in $g(0,x_t)=0$. However, since $g(0,x_t)=-\alpha rR_T\frac{1-\alpha^{N-1}}{1-\alpha}<0$ always holds, there is no extra boundary fixed point when $x_i=0$.

Let $x_i=\alpha$, we restrict $x_t(1-\alpha-x_t)\neq 0$. Solving $\dot{x_i}=0$ and $\dot{x_t}=0$ results in $g(\alpha,x_t)=0$, that is $\lambda \frac{1-(\alpha+x_t)^{N-1}}{1-\alpha-x_t}-rR_T\frac{1-\alpha^{N-1}}{1-\alpha}=0$. Let $\varphi_1(x)=\lambda\frac{1-(\alpha+x)^{N-1}}{1-\alpha-x}-rR_T\frac{1-\alpha^{N-1}}{1-\alpha}=\lambda\sum_{i=0}^{N-2}(\alpha+x)^i-rR_T\frac{1-\alpha^{N-1}}{1-\alpha}$. It is straightforward that $\varphi_1(x)$ increases monotonously from $0$ to $1-\alpha$. Therefore, when $\varphi_1(0)\varphi_1(1-\alpha)<0$, $\varphi_1(x) =0$ has a unique solution $\overline{x}_{t1} \in (0, 1-\alpha)$, and accordingly there is a boundary fixed point $(\alpha,\overline{x}_{t1})$.

Let $x_t=0$, we restrict $x_i(\alpha-x_i)\neq 0$. Solving $\dot{x_i}=0$ and $\dot{x_t}=0$ results in $f(x_i,0)=0 $, that is $ x_i\frac{1-(1-\alpha+x_i)^{N-1}}{\alpha-x_i}-x_i\frac{\alpha^{N-1}-x_i^{N-1}}{\alpha-x_i}+(1-\alpha+x_i)^{N-1}-x_i^{N-1}-2+2\alpha^{N-1}=0$. Let $\varphi_2(x)=x\frac{1-(1-\alpha+x)^{N-1}}{\alpha-x}-x\frac{\alpha^{N-1}-x^{N-1}}{\alpha-x}+(1-\alpha+x)^{N-1}-x^{N-1}-2+2\alpha^{N-1}=x\sum_{i=0}^{N-2}(1-\alpha+x)^i-x\sum_{i=0}^{N-2}\alpha^{N-2-i}x^i+(1-\alpha+x)^{N-1}-x^{N-1}-2+2\alpha^{N-1}$. We have $\varphi_2^{'}(x)=\sum_{i=0}^{N-2}(1-\alpha+x)^{i}+x\sum_{i=1}^{N-2}i(1-\alpha+x)^{i-1}-\sum_{i=0}^{N-2}\alpha^{N-2-i}x^{i}-x\sum_{i=1}^{N-2}i\alpha^{N-2-i}x^{i-1}+(N-1)(1-\alpha+x)^{N-2}-(N-1)x^{N-2}>\sum_{i=0}^{N-2}((1-\alpha+x)^i-x^i)+x\sum_{i=1}^{N-2}i((1-\alpha+x)^{i-1}-x^{i-1})+(N-1)((1-\alpha+x)^{N-2}-x^{N-2})>0$. Since $\varphi_2(x)$ increases monotonously from $0$ to $\alpha$, when $\varphi_2(0)\varphi_2(\alpha)<0$, $\varphi_2(x) =0$ has a unique solution $\overline{x}_{i1} \in (0, \alpha)$, and accordingly there is a boundary fixed point: $(\overline{x}_{i1},0)$.

Let $x_t=1-\alpha$, we restrict $x_i(\alpha-x_i)\neq 0$. Solving $\dot{x_i}=0$ and $\dot{x_t}=0$ results in $f(x_i,1-\alpha)=0$, that is $ x_i\frac{1-(1-\alpha+x_i)^{N-1}}{\alpha-x_i}-x_i\frac{\alpha^{N-1}-x_i^{N-1}}{\alpha-x_i}+(1-\alpha+x_i)^{N-1}-x_i^{N-1}-1+\alpha^{N-1}=0$. Let
$\varphi_3(x)=x\frac{1-(1-\alpha+x)^{N-1}}{\alpha-x}-x\frac{\alpha^{N-1}-x^{N-1}}{\alpha-x}+(1-\alpha+x)^{N-1}-x^{N-1}-1+\alpha^{N-1}$, we have $\varphi_3(x)=\varphi_2(x)+1-\alpha^{N-1}$, so $\varphi_3(x)$ increases monotonously from $0$ to $\alpha$. Since  $\varphi_3(0)\varphi_3(\alpha)<0$ always holds, $\varphi_3(x) =0$ has a unique solution $\overline{x}_{i2} \in (0, \alpha)$, and accordingly there is a boundary fixed point $(\overline{x}_{i2},1-\alpha)$.

\section{Stabilities of boundary fixed points}

In this section, by keeping the notation introduced in Appendix~A, we investigate the stabilities of the mentioned boundary fixed points.

For simplicity, we denote the unique solution of $(N-1)\alpha-(N-2)\alpha^{N-1}=1$ by $\alpha^*$ (in particular, $\alpha^*=1$ for $N=3$). It can be seen that $(N-1)\alpha-(N-2)\alpha^{N-1} < 1$ when $\alpha < \alpha^*$ and $(N-1)\alpha-(N-2)\alpha^{N-1}>1$ when $\alpha > \alpha^*$.

Now, we have the Jacobian matrix of the system
\begin{equation}
	J={\left[ \begin{array}{ccc}
			J_{11}&J_{12}\\
			J_{21}&J_{22}\nonumber
		\end{array}
		\right ]},
\end{equation}
where
\[\begin{cases}
	J_{11}=\frac{\partial \dot{x_i}}{\partial x_i}=\frac{\lambda t_v}{\alpha}((\alpha-2x_i)f(x_i,x_t)+x_i(\alpha-x_i)\frac{\partial f(x_i,x_t)}{\partial x_i}),\\
	J_{12}=\frac{\partial \dot{x_i}}{\partial x_t}=\frac{\lambda t_v}{\alpha}x_i(\alpha-x_i)\frac{\partial f(x_i,x_t)}{\partial x_t}=\frac{\lambda t_v}{\alpha}(N-1)x_i(\alpha-x_i)(\alpha+x_t)^{N-2},\\
	J_{21}=\frac{\partial \dot{x_t}}{\partial x_i}=\frac{\lambda t_v}{1-\alpha}x_t(1-\alpha-x_t)\frac{\partial g(x_i,x_t)}{\partial x_i}=\frac{\lambda t_v}{1-\alpha}x_t(1-(\alpha+x_t)^{N-1}),\\
	J_{22}=\frac{\partial \dot{x_t}}{\partial x_t}=\frac{t_v}{1-\alpha}((1-\alpha -2x_t)g(x_i,x_t)+x_t(1-\alpha-x_t)\frac{\partial g(x_i,x_t)}{\partial x_t})\,.
\end{cases}\]

We have the following formulas
\[\begin{cases}
	\frac{\partial f(x_i,x_t)}{\partial x_i}=\frac{\partial \varphi_2(x_i)}{\partial x_i}=\varphi_2^{'}(x_i)>0,\\
	\frac{\partial g(x_i,x_t)}{\partial x_t}=x_i\frac{\partial \varphi_1(x_t)}{\partial x_t}=x_i\varphi_1^{'}(x_t)>0  \text{ for }\ x_i \in (0,\alpha).
\end{cases}\]

Next, we analyze the stabilities of boundary fixed points in different parameter regions.

\subsection{For $\alpha < \alpha^*$ and $ \lambda<rR_T(1-\alpha^{N-1})/((N-1)(1-\alpha))$}
$\varphi_1(0)\varphi_1(1-\alpha)<0$ is equivalent to $rR_T(1-\alpha^{N-1})/((N-1)(1-\alpha))<\lambda<rR_T$, and $\varphi_2(0)\varphi_2(\alpha)<0$ is equivalent to $(N-1) \alpha-(N-2) \alpha^{N-1}>1$. Thus, when $\alpha < \alpha^*$ (i.e., $(N-1) \alpha-(N-2) \alpha^{N-1}<1$) and $ \lambda<rR_T(1-\alpha^{N-1})/((N-1)(1-\alpha))$, the boundary fixed points $(\alpha,\overline{x}_{t1})$ and $(\overline{x}_{i1},0)$ do not exist, hence there are the following five boundary fixed points: $(0,0)$, $(0,1-\alpha)$, $(\alpha,0)$, $(\alpha,1-\alpha)$, and $(\overline{x}_{i2},1-\alpha)$.

For the boundary fixed point $(0,0)$, the Jacobian matrix is
\begin{equation}
	J={\left[ \begin{array}{ccc}
			J_{11}&0\\
			0&J_{22}\nonumber
		\end{array}
		\right ]}.
\end{equation}
Since $J_{11}=\lambda t_v((1-\alpha)^{N-1}+2\alpha^{N-1}-2)=\lambda t_v((\alpha^{N-1}+(1-\alpha)^{N-1}-1)+(\alpha^{N-1}-1))<0$, $J_{22}=-\alpha rR_Tt_v\frac{1-\alpha^{N-1}}{1-\alpha}<0$, $(0,0)$ is always stable.

For the boundary fixed point $(0,1-\alpha)$, the Jacobian matrix  is
\begin{equation}
	J={\left[ \begin{array}{ccc}
			J_{11}&0\\
			0&J_{22}\nonumber
		\end{array}
		\right ]}.
\end{equation}
Since $J_{22}=\alpha rR_Tt_v\frac{1-\alpha^{N-1}}{1-\alpha}>0$, $(0,1-\alpha)$ is always unstable.

For the boundary fixed point $(\alpha,0)$, the Jacobian matrix  is
\begin{equation}
	J={\left[ \begin{array}{ccc}
			J_{11}&0\\
			0&J_{22}\nonumber
		\end{array}
		\right ]}.
\end{equation}
Since $J_{11}=t_v(1+(N-2)\alpha^{N-1}-(N-1)\alpha)>0$, $(\alpha,0)$ is unstable.

For the boundary fixed point $(\alpha,1-\alpha)$, the Jacobian matrix  is
\begin{equation}
	J={\left[ \begin{array}{ccc}
			J_{11}&0\\
			0&J_{22}\nonumber
		\end{array}
		\right ]}.
\end{equation}
Since $J_{22}=-\alpha t_v(\lambda(N-1)- rR_T\frac{ 1-\alpha^{N-1}}{1-\alpha})>0$,  $(\alpha,1-\alpha)$ is unstable.

For the boundary fixed point $(\overline{x}_{i2},1-\alpha)$, the Jacobian matrix  is
\begin{equation}
	J={\left[ \begin{array}{ccc}
			J_{11}&J_{12}\\
			0&J_{22}\nonumber
		\end{array}
		\right ]}.
\end{equation}
Since $J_{11}=\frac{\lambda t_v}{\alpha}\overline{x}_{i2}(\alpha-\overline{x}_{i2})\left.\frac{\partial f(x_i,1-\alpha)}{\partial x_i}\right|_{x_i=\overline{x}_{i2}}>0$,  $(\overline{x}_{i2},1-\alpha)$ is always unstable.

In sum, among all these five boundary fixed points, only $(0,0)$ is stable.

\subsection{For $\alpha > \alpha^*$ and $ \lambda<rR_T(1-\alpha^{N-1})/((N-1)(1-\alpha))$}
$\varphi_1(0)\varphi_1(1-\alpha)<0$ is equivalent to $rR_T(1-\alpha^{N-1})/((N-1)(1-\alpha))<\lambda<rR_T$, and $\varphi_2(0)\varphi_2(\alpha)<0$ is equivalent to $(N-1) \alpha-(N-2) \alpha^{N-1}>1$. Thus, when $\alpha > \alpha^*$ (i.e., $(N-1) \alpha-(N-2) \alpha^{N-1}>1$) and $ \lambda<rR_T(1-\alpha^{N-1})/((N-1)(1-\alpha))$, the boundary fixed point $(\alpha,\overline{x}_{t1})$ does not exist, there are six boundary fixed points: $(0,0)$, $(0,1-\alpha)$, $(\alpha,0)$, $(\alpha,1-\alpha)$, $(\overline{x}_{i1},0)$, and $(\overline{x}_{i2},1-\alpha)$. As stated in Appendix~B.1, the boundary fixed point $(0,0)$ is stable, while both the boundary fixed points $(0,1-\alpha)$ and $(\overline{x}_{i2},1-\alpha)$ are  unstable.

For the boundary fixed point $(\alpha,0)$, the Jacobian matrix  is
\begin{equation}
	J={\left[ \begin{array}{ccc}
			J_{11}&0\\
			0&J_{22}\nonumber
		\end{array}
		\right ]}.
\end{equation}
Since $J_{11}=t_v(1+(N-2)\alpha^{N-1}-(N-1)\alpha)<0$, $J_{22}=\alpha t_v(\lambda-rR_T)\frac{1-\alpha^{N-1}}{1-\alpha}<0$, $(\alpha,0)$ is stable.

For the boundary fixed point $(\alpha,1-\alpha)$, the Jacobian matrix  is
\begin{equation}
	J={\left[ \begin{array}{ccc}
			J_{11}&0\\
			0&J_{22}\nonumber
		\end{array}
		\right ]}.
\end{equation}
Since $J_{22}=-\alpha t_v(\lambda(N-1)- rR_T\frac{ 1-\alpha^{N-1}}{1-\alpha})>0$,  $(\alpha,1-\alpha)$ is unstable.

For the boundary fixed point $(\overline{x}_{i1},0)$, the Jacobian matrix  is
\begin{equation}
	J={\left[ \begin{array}{ccc}
			J_{11}&J_{12}\\
			0&J_{22}\nonumber
		\end{array}
		\right ]}.
\end{equation}
Since $J_{11}=\frac{\lambda t_v}{\alpha}\overline{x}_{i1}(\alpha-\overline{x}_{i1})\left.\frac{\partial f(x_i,0)}{\partial x_i}\right|_{x_i=\overline{x}_{i1}}>0$,  $(\overline{x}_{i1},0)$ is always unstable.

Altogether, among all these six boundary fixed points, $(0,0)$ and $(\alpha,0)$ are stable.

\subsection{For $\alpha < \alpha^*$ and $rR_T(1-\alpha^{N-1})/((N-1)(1-\alpha))<\lambda<rR_T$}
$\varphi_1(0)\varphi_1(1-\alpha)<0$ is equivalent to $rR_T(1-\alpha^{N-1})/((N-1)(1-\alpha))<\lambda<rR_T$, and $\varphi_2(0)\varphi_2(\alpha)<0$ is equivalent to $(N-1) \alpha-(N-2) \alpha^{N-1}>1$. Thus, when $\alpha < \alpha^*$ (i.e., $(N-1) \alpha-(N-2) \alpha^{N-1}<1$) and  $rR_T(1-\alpha^{N-1})/((N-1)(1-\alpha))<\lambda<rR_T$, the boundary fixed point $(\overline{x}_{i1},0)$ does not exist, there are six boundary fixed points: $(0,0)$, $(0,1-\alpha)$, $(\alpha,0)$, $(\alpha,1-\alpha)$, $(\alpha,\overline{x}_{t1})$, and $(\overline{x}_{i2},1-\alpha)$. As stated in Appendix~B.1, the boundary fixed point $(0,0)$ is stable, while both the boundary fixed points $(0,1-\alpha)$ and $(\overline{x}_{i2},1-\alpha)$ are  unstable.

For the boundary fixed point $(\alpha,0)$, the Jacobian matrix  is
\begin{equation}
	J={\left[ \begin{array}{ccc}
			J_{11}&0\\
			0&J_{22}\nonumber
		\end{array}
		\right ]}.
\end{equation}
Since $J_{11}=t_v(1+(N-2)\alpha^{N-1}-(N-1)\alpha)>0$, $(\alpha,0)$ is unstable.

For the boundary fixed point $(\alpha,1-\alpha)$, the Jacobian matrix  is
\begin{equation}
	J={\left[ \begin{array}{ccc}
			J_{11}&0\\
			0&J_{22}\nonumber
		\end{array}
		\right ]}.
\end{equation}
Since $J_{11}=-\lambda t_v((N-1)\alpha-(N-1)\alpha^{N-1})<0$, $J_{22}=-\alpha t_v(\lambda(N-1)- rR_T\frac{ 1-\alpha^{N-1}}{1-\alpha})<0$,  $(\alpha,1-\alpha)$ is stable.

For the boundary fixed point $(\alpha,\overline{x}_{t1})$, the Jacobian matrix  is
\begin{equation}
	J={\left[ \begin{array}{ccc}
			J_{11}&0\\
			J_{21}&J_{22}\nonumber
		\end{array}
		\right ]}.
\end{equation}
Since $J_{22}=\frac{t_v}{1-\alpha}\overline{x}_{t1}(1-\alpha-\overline{x}_{t1})\left.\frac{\partial g(\alpha,x_t)}{\partial x_t} \right|_{x_t=\overline{x}_{t1}}\ >0$,  $(\alpha,\overline{x}_{t1})$ is always unstable.

Therefore, among all these six boundary fixed points, $(0,0)$ and $(\alpha,1-\alpha)$ are stable.

\subsection{For $\alpha > \alpha^*$ and $rR_T(1-\alpha^{N-1})/((N-1)(1-\alpha))<\lambda<rR_T$}
$\varphi_1(0)\varphi_1(1-\alpha)<0$ is equivalent to $rR_T(1-\alpha^{N-1})/((N-1)(1-\alpha))<\lambda<rR_T$, and $\varphi_2(0)\varphi_2(\alpha)<0$ is equivalent to $(N-1) \alpha-(N-2) \alpha^{N-1}>1$. Thus, when $\alpha > \alpha^*$ (i.e., $(N-1) \alpha-(N-2) \alpha^{N-1}>1$) and $rR_T(1-\alpha^{N-1})/((N-1)(1-\alpha))<\lambda<rR_T$, there are seven boundary fixed points: $(0,0)$, $(0,1-\alpha)$, $(\alpha,0)$, $(\alpha,1-\alpha)$, $(\alpha,\overline{x}_{t1})$, $(\overline{x}_{i1},0)$, and $(\overline{x}_{i2},1-\alpha)$. As stated in Appendix~B.2 and Appendix~B.3, the boundary fixed point $(0,0)$ is stable, while the boundary fixed points $(0,1-\alpha)$, $(\alpha,\overline{x}_{t1})$, $(\overline{x}_{i1},0)$, and $(\overline{x}_{i2},1-\alpha)$ are all unstable.

For the boundary fixed point $(\alpha,0)$, the Jacobian matrix  is
\begin{equation}
	J={\left[ \begin{array}{ccc}
			J_{11}&0\\
			0&J_{22}\nonumber
		\end{array}
		\right ]}.
\end{equation}
Since $J_{11}=t_v(1+(N-2)\alpha^{N-1}-(N-1)\alpha)<0$, $J_{22}=\alpha t_v(\lambda-rR_T)\frac{1-\alpha^{N-1}}{1-\alpha}<0$, $(\alpha,0)$ is stable.

For the boundary fixed point $(\alpha,1-\alpha)$, the Jacobian matrix  is
\begin{equation}
	J={\left[ \begin{array}{ccc}
			J_{11}&0\\
			0&J_{22}\nonumber
		\end{array}
		\right ]}.
\end{equation}
Since $J_{11}=-\lambda t_v((N-1)\alpha-(N-1)\alpha^{N-1})<0$, $J_{22}=-\alpha t_v(\lambda(N-1)- rR_T\frac{ 1-\alpha^{N-1}}{1-\alpha})<0$,  $(\alpha,1-\alpha)$ is stable.

Among all these seven boundary fixed points, $(0,0)$, $(\alpha,0)$, and $(\alpha,1-\alpha)$ are stable.

\subsection{For $\alpha \leq \alpha^*$ and $\lambda>rR_T$}
$\varphi_1(0)\varphi_1(1-\alpha)<0$ is equivalent to $rR_T(1-\alpha^{N-1})/((N-1)(1-\alpha))<\lambda<rR_T$, and $\varphi_2(0)\varphi_2(\alpha)<0$ is equivalent to $(N-1) \alpha-(N-2) \alpha^{N-1}>1$. Thus, when $\alpha \leq \alpha^*$ (i.e., $(N-1) \alpha-(N-2) \alpha^{N-1} \leq 1$) and $\lambda>rR_T$, the boundary fixed points $(\alpha,\overline{x}_{t1})$ and $(\overline{x}_{i1},0)$ do not exist, there are five boundary fixed points: $(0,0)$, $(0,1-\alpha)$, $(\alpha,0)$, $(\alpha,1-\alpha)$, and $(\overline{x}_{i2},1-\alpha)$. As stated in Appendix~B.1, the boundary fixed point $(0,0)$ is stable, while both the boundary fixed points $(0,1-\alpha)$ and $(\overline{x}_{i2},1-\alpha)$ are unstable.

For the boundary fixed point $(\alpha,0)$, the Jacobian matrix  is
\begin{equation}
	J={\left[ \begin{array}{ccc}
			J_{11}&0\\
			0&J_{22}\nonumber
		\end{array}
		\right ]}.
\end{equation}
Since $J_{22}=\alpha t_v(\lambda-rR_T)\frac{1-\alpha^{N-1}}{1-\alpha}>0$, $(\alpha,0)$ is unstable.

For the boundary fixed point $(\alpha,1-\alpha)$, the Jacobian matrix  is
\begin{equation}
	J={\left[ \begin{array}{ccc}
			J_{11}&0\\
			0&J_{22}\nonumber
		\end{array}
		\right ]}.
\end{equation}
Since $J_{11}=-\lambda t_v((N-1)\alpha-(N-1)\alpha^{N-1})<0$, $J_{22}=-\alpha t_v(\lambda(N-1)-rR_T \frac{1-\alpha^{N-1}}{1-\alpha})<0$,  $(\alpha,1-\alpha)$ is stable.

Among all these five boundary fixed points, $(0,0)$ and $(\alpha,1-\alpha)$ are stable.

\subsection{For $\alpha > \alpha^*$ and $\lambda>rR_T$}
$\varphi_1(0)\varphi_1(1-\alpha)<0$ is equivalent to $rR_T(1-\alpha^{N-1})/((N-1)(1-\alpha))<\lambda<rR_T$, and $\varphi_2(0)\varphi_2(\alpha)<0$ is equivalent to $(N-1) \alpha-(N-2) \alpha^{N-1}>1$. Thus, when  $\alpha > \alpha^*$ (i.e., $(N-1) \alpha-(N-2) \alpha^{N-1}>1$) and $\lambda>rR_T$, the boundary fixed point $(\alpha,\overline{x}_{t1})$ does not exist, there are six boundary fixed points: $(0,0)$, $(0,1-\alpha)$, $(\alpha,0)$, $(\alpha,1-\alpha)$, $(\overline{x}_{i1},0)$, and $(\overline{x}_{i2},1-\alpha)$. As stated in Appendix~B.2, the boundary fixed point $(0,0)$ is stable, while the boundary fixed points $(0,1-\alpha)$,  $(\overline{x}_{i1},0)$, and $(\overline{x}_{i2},1-\alpha)$ are all unstable.

For the boundary fixed point $(\alpha,0)$, the Jacobian matrix  is
\begin{equation}
	J={\left[ \begin{array}{ccc}
			J_{11}&0\\
			0&J_{22}\nonumber
		\end{array}
		\right ]}.
\end{equation}
Since $J_{22}=\alpha t_v(\lambda-rR_T)\frac{1-\alpha^{N-1}}{1-\alpha}>0$, $(\alpha,0)$ is unstable.

For the boundary fixed point $(\alpha,1-\alpha)$, the Jacobian matrix  is
\begin{equation}
	J={\left[ \begin{array}{ccc}
			J_{11}&0\\
			0&J_{22}\nonumber
		\end{array}
		\right ]}.
\end{equation}
Since $J_{11}=-\lambda t_v((N-1)\alpha-(N-1)\alpha^{N-1})<0$, $J_{22}=-\alpha t_v(\lambda(N-1)-rR_T \frac{1-\alpha^{N-1}}{1-\alpha})<0$, $(\alpha,1-\alpha)$ is stable.

Among all these six boundary fixed points, $(0,0)$ and $(\alpha,1-\alpha)$ are stable.

\section{Stabilities of interior fixed points}
In this section, following the notation used in Appendix~A and Appendix~B, we investigate the stabilities of interior fixed points if they exist.

Since the replicator equations are highly non-linear, it is difficult to prove whether there is an interior fixed point. If there exists an interior fixed point, we then formally write it as $(\overline{x}_{i3},\overline{x}_{t3})$, where $\overline{x}_{i3} \in (0,\alpha)$ and $\overline{x}_{t3} \in (0,1-\alpha)$.

Accordingly, we have
\begin{equation}
	\left\{
	\begin{array}{lr}
		\dot{x_i}=\frac{1}{\alpha}\lambda t_vx_i(\alpha-x_i)f(x_i,x_t),\\
		\dot{x_t}=\frac{1}{1-\alpha}t_vx_t(1-\alpha-x_t)g(x_i,x_t).
	\end{array}\nonumber
	\right.
\end{equation}

When $x_i(\alpha-x_i)\neq 0$ and $x_t(1-\alpha-x_t)\neq 0$, solving $\dot{x_i}=0$ and $\dot{x_t}=0$ results in
\[\begin{cases}
	f(x_i,x_t)=0,\\
	g(x_i,x_t)=0.
\end{cases}\]
That is, both $f(\overline{x}_{i3},\overline{x}_{t3})=0$ and $g(\overline{x}_{i3},\overline{x}_{t3})=0$ are satisfied.

Note that the Jacobian matrix  is
\begin{equation}
	J={\left[ \begin{array}{ccc}
			J_{11}&J_{12}\\
			J_{21}&J_{22}\nonumber
		\end{array}
		\right ]}.
\end{equation}

As stated in Appendix~B, we have
\[\begin{cases}
	J_{11}=\frac{\lambda t_v}{\alpha}\overline{x}_{i3}(\alpha-\overline{x}_{i3})\left.\frac{\partial f(x_i,\overline{x}_{t3})}{\partial x_i}\right|_{x_i=\overline{x}_{i3}}>0,\\
	J_{12}=\frac{\lambda t_v}{\alpha}(N-1)\overline{x}_{i3}(\alpha-\overline{x}_{i3})(\alpha+\overline{x}_{t3})^{N-2}>0,\\
	J_{21}=\frac{\lambda t_v}{1-\alpha}\overline{x}_{t3}(1-(\alpha+\overline{x}_{t3})^{N-1})>0,\\
	J_{22}=\frac{ t_v}{1-\alpha}\overline{x}_{t3}(1-\alpha-\overline{x}_{t3})\left.\frac{\partial g(\overline{x}_{i3},x_t)}{\partial x_t}\right|_{x_t=\overline{x}_{t3}}>0.
\end{cases}\]

Let
\begin{equation}
	\phi(\lambda)={\left| \begin{array}{ccc}
			\lambda-J_{11} & &-J_{12}\\
			-J_{21} & & \lambda -J_{22}\nonumber
		\end{array}
		\right |}.
\end{equation}

We can see that ${ \lim_{\lambda\to +\infty} \phi(\lambda) = +\infty}$, and $\phi(J_{11})=-J_{12}J_{21}<0$.
Since $J_{11}>0$ and $\phi(\lambda)$ is a continuous function of $\lambda$, $\phi(\lambda)=0$ has at least a positive solution, that is, the Jacobian matrix has at least a positive eigenvalue. Thus, the interior fixed point $(\overline{x}_{i3},\overline{x}_{t3})$ is unstable.
\end{document}